\documentclass[aps,prd,floatfix,superscriptaddress,showpacs]{revtex4}

\usepackage{epsfig}
\usepackage{amsmath}
\usepackage{graphicx,psfrag}
\usepackage{dcolumn}
\usepackage{bm}
\usepackage{slashed}
\usepackage{color}

\newcommand{\beq}{\begin{equation}}
\newcommand{\eeq}{\end{equation}}
\newcommand{\bea}{\begin{eqnarray}}
\newcommand{\eea}{\end{eqnarray}}
\newcommand{\hf} {\frac{1}{2}}

\newcommand{\nn}{\nonumber\\}

\newcommand\eqn[1]     {Eq.\,(\ref{#1})}
\newcommand\fig[1]     {Fig.\,{\ref{#1}}}
\newcommand\app[1]     {Appendix~\ref{#1}}

\def\Tr{{\rm Tr}}

\def\eq#1{(\ref{#1})}
\def\s0#1#2{\mbox{\small{$ \frac{#1}{#2} $}}}
\def\0#1#2{\frac{#1}{#2}}

\def\ord#1{{\cal O}(#1)}

\def\mr#1{{\mathrm{#1}}}

\begin{document}

\title{Coulomb Gas and Sine-Gordon Model in Arbitrary Dimension}

\author{I. N\'andori}
\affiliation{University of Debrecen, Institute of Physics, H-4010 Debrecen P.O. Box 105, Hungary}
\affiliation{MTA-DE Particle Physics Research Group, P.O.Box 51, H-4001 Debrecen, Hungary}
\affiliation{Institute of Nuclear Research, P.O.Box 51, H-4001 Debrecen, Hungary} 

\begin{abstract} 
The sine-Gordon (SG), i.e. periodic scalar field theory is known to play an important role in $d=2$ 
dimensions. A paradigmatic example is the topological phase transition of the vortex dynamics in 
superfluid films and layered superconductors which are described by SG type models. Periodic 
scalar potentials find applications in $d=4$ dimensions, too. Higgs, inflaton and axion physics are 
examples where scalar fields naturally appear, thus, the SG model can be used instead of the usual 
polynomial one. The SG quantum field theory can be mapped onto the neutral Coulomb-gas (CG) 
in arbitrary dimension and the renormalization group (RG) study of the d-dimensional CG model 
was obtained in the dilute gas approximation. It signals a single phase for $d>2$, however, it was 
shown recently, that a suitable generalization of the SG model can posses a topological 
phase transitions in $d=4$ dimensions. Our goals in this work are (i) to map out the phase structure 
of the (original) SG and the equivalent neutral CG models by the functional RG method 
in arbitrary dimension, (ii) to compare the 3-dimensional SG and isotropic XY spin models and 
show that they belong to different universality classes, (iii) to study the consequences of the findings 
on higgs, inflaton, axion models and on the topological phase transition in higher dimensions.
\end{abstract}


\maketitle

\section{Introduction}
\label{sec_intro}

Universality expresses the idea that different microscopic physics can give rise to the same scaling 
behavior at a phase transition. If the  scaling behaviour of different models agrees or disagrees it is 
possible to decide whether they belong to the same universality class or not. If two different models 
can be mapped onto each other they belong to the same universality class hence by the study of one 
model one can obtain the scaling behaviour of the other one.  Furthermore, models of the same class 
of universality can be used to test and compare various types of methods. A canonical example for 
model equivalences is the sine-Gordon (SG) scalar field theory defined by the Euclidean action 
\cite{zj,trunc_rg_sg,d3_sg,d3_sg_direct,d4_sg}
\beq
\label{sg}
S = \int d^d x \left[\hf (\partial_\mu\varphi_x)^2 + u\cos(\beta \varphi_x)\right],
\eeq
which belongs to the universality class of the neutral Coulomb gas (CG) \cite{samuel} and the isotropic XY 
classical spin model \cite{zj} for $d=2$. They undergo a topological, i.e., Kosterlitz-Thouless-Berezinski 
\cite{KTB} (KTB) phase transition. Moreover, the KTB universality regulates the scaling 
behaviour of the O(N) symmetric field theory with $d=N=2$. It is worth noting that signatures of the KTB 
scaling in this model have been also observed by the Functional Renormalization Group (FRG) method 
\cite{d2_o2}.
Furthermore, the SG scalar theory is the bose form of the 2-dimensional fermionic Thirring model 
\cite{coleman} and the bosonic counterpart of multi-flavor Quantum Electrodynamics and multi-color 
Quantum Chromodynamics are also SG type models \cite{qed_qcd}. Bosonization is well established only 
in $d=2$ dimensions but the mapping between the SG theory \eq{sg} and the neutral CG holds 
in arbitrary dimension \cite{zj,d3_sg}. 

In $d=2$ dimensions, the SG model has been used to describe the vortex dynamics in superfluid films 
and magnetically coupled layered superconductors \cite{layered_sg}. In addition, the holographic RG flow 
\cite{holographic_rg} and the the c-function \cite{c_func_sg} of the SG model have also been studied. 

In $d>2$ dimensions one might expect no room for any physical application for the SG model, however, 
there are various cases where periodic scalar fields can play a role in 4-dimensional physics. The most 
natural three situations among these are the following (i) the periodic inflationary potential, (ii) the mass 
generation by a periodic Higgs potential, (iii) the axion potential which naturally appears as a periodic 
function. The main concepts of these three cases are summarised in the appendices. For example, in 
inflationary cosmology the so called natural inflation i.e., a periodic potential has already been used as 
a competing inflationary model, see \app{sg_inflaton}. Another possible application is a periodic 
self-interaction which is proposed here as a possible extension of the standard model Higgs potential. 
Using the usual parametrisation of the field around the first minimum, one recovers the Lagrangian 
of the Higgs sector but with a periodic interaction term for a single-component real scalar field, see 
\app{periodic_higgs}. Finally, one has to mention the periodic axion potential which was proposed 
to retain the CP conserving nature of QCD where gauge symmetry and renormalizability allow the 
inclusion of CP violating terms but experimental data do not favour such an extension, see \app{axion}.

Recently \cite{extended_d4_sg}, an extension of the SG scalar theory has been studied in $d=4$ 
dimensions. Eq.(2) of \cite{extended_d4_sg} has an unusual kinetic term $(\Delta \varphi_x)^2$ 
where $\Delta$ is the 4-dimensional Laplacian and $\varphi$ is a real scalar field. On the one hand, 
it was shown in \cite{extended_d4_sg}  by the exact FRG study performed in the momentum space 
that such extension of the SG model undergoes a topological phase transitions. On the other hand,
it is also known that the neutral Coulomb-gas which is identical to the SG model \eq{sg} has a single 
phase in $d>2$ dimensions, at least this is the result of the real space renormalization group (RG) 
study obtained in the so called dilute gas approximation \cite{d_cg}. 
In the real space RG approach the RG transformations are performed in the coordinate 
space by using a sharp cutoff, i.e., the charges of the Coulomb gas are represented by solid discs 
with finite diameter and in the dilute gas approximation, only two-body interactions are taken
into account. This approximation is valid if the fugacity of the Coulomb gas remains small. Since 
the fugacity of the Coulomb gas is directly related to the Fourier amplitude of the SG model, 
the dilute gas approximation corresponds to the case where the Fourier amplitude is assumed to 
be small, so the RG equations can be linearised with respect to the amplitude which gives the 
linearised RG flow. The single phase picture is based on these dilute gas or linearised RG equations.
Thus, it is a relevant question whether the exact phase structure of the SG model 
\eq{sg} posses any topological phase transitions in higher dimensions.
 
Indeed, the RG study of the phase structure of the d-dimensional SG model \eq{sg} has already 
performed in \cite{d3_sg,d3_sg_direct,trunc_rg_sg,d4_sg} using various approximations. For example, 
in Refs.~\cite{d3_sg,d3_sg_direct,d4_sg} the RG equations were obtained in the leading order of the 
derivative expansion of the action, i.e., in the local potential approximation (LPA) where the wave
function renormalization is kept to be constant. In Eq.~\eq{sg}, the scalar field can be rescaled 
($\varphi \to \beta \varphi$) and as a result, the frequency parameter $\beta$ appears in the kinetic 
term and related to the wave function renormalization ($z \equiv 1/\beta^2$). Thus, LPA means 
constant dimensionful frequency. Another type of approximation has been used in \cite{trunc_rg_sg},
where RG equations were obtained beyond LPA but they are expanded in terms of the Fourier 
amplitude. Let us note, that the leading order in this approach is equivalent to the dilute gas 
approximation of the real space RG approach of the corresponding neutral Coulomb-gas \cite{d3_sg}.

The goals of this work are (i) to go beyond the previously used approximations and to map out the exact 
phase structure of the d-dimensional SG model in the framework of the functional RG method, 
(ii) to consider possible equivalences between the SG scalar field theory, the neutral CG and the XY spin 
models in $d\neq 2$ dimensions, (iii) to study the appearance of the topological phase transition in higher 
dimensions, (iv) to draw some conclusions on phase structure of a periodic Higgs, inflaton and axion potentials. 

First the functional RG study of the single and double frequency SG model is presented and the phase 
structures of the SG, CG and XY models are compared. The findings of the present work is compared 
to recent results, then applied for Higgs, inflaton and axion physics. Finally results are summarised in the 
conclusion.

\section{Functional RG study of the SG model}
\label{sec_sg}
The main purpose here is to perform the functional RG study of the SG model in arbitrary dimension (in particular 
in $d=4$) going beyond approximations used in previous works \cite{d3_sg,d3_sg_direct,trunc_rg_sg,d4_sg}. 
This is achieved by extending the RG analysis of the single-frequency two-dimensional SG model \cite{ncut1} 
for higher dimensions and for higher harmonics. There is an interest in the literature to improve the RG study of 
the d-dimensional periodic model (or the equivalent Coulomb gas), e.g., the PhD thesis \cite{barkhudarov} was 
devoted to the RG study of the Coulomb gas (in the dilute gas approximation which is equivalent to the linearised 
RG flow) for arbitrary dimensions. Findings of this work recover these "dilute gas results" and go beyond that 
by studying the full RG flow equations including wave function renormalization and including higher harmonics 
in arbitrary dimension. 

We use the effective average action FRG method \cite{eea_rg,Mo1994} where the evolution equation reads as 
\beq
\label{feveq}
k\partial_k\Gamma_k[\varphi] = \hf\mr{Tr}  \left[ \frac{k\partial_k R_k}{R_k+\delta^2_{\varphi}\Gamma_k[\varphi]}\right] \, ,
\eeq
with the  running momentum cutoff $k$ and the Tr stands for the integration over all momenta and $R_k$ is the 
regulator function specified later. In order to solve the FRG equation, we consider the following ansatz for the 
SG model
\beq
\label{eaans_dimful}
\Gamma_k = \int d^d x  \left[\hf z_k (\partial_\mu{\varphi}_x)^2 + V_k({\varphi}_x) \right], 
\hskip 0.3cm V_k(\varphi_x) = u_k \cos(\beta \varphi_x)
\eeq
where the local potential contains a single Fourier mode. The Fourier amplitude $u_k$ and the field independent 
wavefunction renormalization $z_k$ depend on the RG scale $k$. The dimensionful frequency $\beta$, is scale 
independent, i.e., it remains constant over the RG flow because the RG transformation retains the periodicity of 
the dimensionful model with an unchanged period length. Thus, $\beta$ is a free parameter of the model which 
can be chosen arbitrarily. Although RG transformations generate higher harmonics, 
we use the ansatz \eq{eaans_dimful} which contains a single Fourier mode since in case of the 2-dimensional SG 
model \cite{d2_sg} it was found to be an appropriate approximation \cite{ncut1,sg_as}. (Later the role of higher 
harmonics is investigated.) \eqn{feveq_dimless} leads to the RG evolution equations for the blocked potential 
$V_k(\varphi)$ and the wavefunction renormalization $z_k$, 
\begin{align}
\label{ea_v}
&k\partial_k V_k = \hf  \int \frac{d^d p}{(2\pi)^d}  {\cal D}_k \, k\partial_k R_k,\\
\label{ea_z}
&k\partial_k z_k = \left(\frac{\beta}{2\pi} \int_{0}^{2\pi/\beta} d\varphi\right) (V_k''')^2 \int \frac{d^d p}{(2\pi)^d} {\cal D}_k^2 \, k\partial_k R_k 
\left(\frac{2}{d}\frac{\partial^2{\cal D}_k}{\partial p^2\partial p^2}p^2 +\frac{\partial{\cal D}_k}{\partial p^2} \right),
\end{align}
with ${\cal D}_k = (z_k p^2 + R_k + V''_k)^{-1}$ where $V''_k \equiv \partial^2_\varphi V_k$ and $V'''_k \equiv \partial^3_\varphi V_k$. 
Since the l.h.s of \eq{ea_z} is independent of the field, a projection onto the 
field-independent subspace has been introduced on the r.h.s of  \eq{ea_z}. The scale $k$ covers the momentum interval from the 
high-energy/ultraviolet (UV) cutoff $\Lambda$ to zero. Inserting the ansatz \eq{eaans_dimful} into Eqs. \eq{ea_v}, \eq{ea_z}, flow 
equations for the dimensionful couplings can be derived
\bea
\label{exact_u}
k\partial_k u_k &=&
\int _p \frac{k\partial_k R_k}{\beta^2 u_k}
\left(\frac{P-\sqrt{P^2-(\beta^2 u_k)^2}}{\sqrt{P^2-(\beta^2 u_k)^2}}\right),\\
\label{exact_z}
k\partial_k z_k &=& \int_p \frac{\beta^2 \, k\partial_k R_k}{2}
\biggl[
\frac{-(\beta^2 u_k)^2P(\partial_{p^2}P+\frac{2}{d}p^2\partial_{p^2}^2P)}
{[P^2-(\beta^2u_k)^2]^{5/2}} 
+\frac{(\beta^2 u_k)^2 p^2 (\partial_{p^2}P)^2(4P^2+(\beta^2 u_k)^2)}
{d \, [P^2-(\beta^2 u_k)^2]^{7/2}}
\biggr] ,
\eea
where $P = z_k p^2+R_k$ and $\int_p = \int dp \, p^{d-1} \Omega_d/(2\pi)^d$  with the d-dimensional solid angle 
$\Omega_d$. Since the dimensionful frequency is scale-independent, it is convenient to merge it with the scale-dependent 
wave function renormalization $z_k$. Thus, we introduce $\hat z_k = z_k/\beta^2$, $\hat R_k = R_k/\beta^2$ and 
$\hat P =  P/\beta^2 = \hat z_k p^2+ \hat R_k$ and the RG flow equations \eq{exact_u} and \eq{exact_z} can be written as,
\bea
\label{exact_u_scaled}
k\partial_k u_k &=&
\int _p \frac{k\partial_k \hat R_k}{u_k}
\left(\frac{\hat P-\sqrt{\hat P^2-u_k^2}}{\sqrt{\hat P^2-u_k^2}}\right),\\
\label{exact_z_scaled}
k\partial_k \hat z_k &=& \int_p \frac{k\partial_k \hat R_k}{2}
\biggl[
\frac{-u_k^2 \hat P(\partial_{p^2} \hat P+\frac{2}{d}p^2\partial_{p^2}^2 \hat P)}
{[\hat P^2-u_k^2]^{5/2}} 
+\frac{u_k^2 p^2 (\partial_{p^2} \hat P)^2(4 \hat P^2+u_k^2)}
{d \, [\hat P^2- u_k^2]^{7/2}}
\biggr] ,
\eea
Before we further study the RG flow equations, we show that they can be derived by using the rescaled version of the 
original action for the SG model,
\beq
\label{eaans_dimless}
\Gamma_k[\theta] = \int d^d x  \left[\hf \hat{z}_k (\partial_\mu{\theta}_x)^2 +  u_k \cos(\theta_x) \right], 
\eeq
where the rescaled (dimensionless) field $\theta = \beta \varphi$ is introduced. Let us note, that the field carries a 
dimension for $d\neq 2$ thus the frequency of the SG model \eq{sg} becomes a dimensionful parameter for $d\neq 2$, 
i.e.,  $\beta^2 = k^{2-d} \tilde\beta_k^2$ where $\tilde \beta_k$ is dimensionless, so the rescaled wavefunction renormalization
has a dimension of $k^{d-2}$, i.e., $\hat z_k = \tilde z_k k^{d-2}$. The corresponding FRG equation reads as
\beq
\label{feveq_dimless}
k\partial_k \Gamma_{k}[\theta]  =  \hf \mr{Tr}  \left[\frac{k\partial_k R_k}{R_k+\beta^2 \delta^2_{\theta} \Gamma_k[\theta]}\right]
=  \hf \mr{Tr}  \left[\frac{k\partial_k \hat R_k}{\hat R_k + \delta^2_{\theta} \Gamma_k[\theta]}\right]
\eeq
where the rescaled regulator function $\hat R_k = R_k/\beta^2$ has been used. Inserting the ansatz \eq{eaans_dimless} 
into Eqs.~\eq{feveq_dimless}, the RG flow equations \eq{exact_u_scaled} and \eq{exact_z_scaled} can be obtained
Thus, equations \eq{exact_u_scaled} and \eq{exact_z_scaled} are derived in two different ways. 

Momentum integrals have to be performed numerically, except the linearized form of Eqs.~\eq{exact_u_scaled}, \eq{exact_z_scaled} 
around the Gaussian fixed point where analytical results available. This requires a special choice for the regulator function $R_k$
such as the power-law \cite{Mo1994} or the Litim-type \cite{opt_rg} ones, 
\beq
R_{\mr{pow}}(k) = p^2 \left(\frac{k^2}{p^2}\right)^b, \hskip 0.5cm R_{\mr{Litim}}(k) = p^2 \left(\frac{k^2}{p^2}-1\right) \Theta(k^2 - p^2)
\eeq
where $\Theta(x)$ stands for the Heaviside step function and $b$ is a free parameter of the power-law regulator.
In general, the regulator function beyond LPA should be given by the inclusion (multiplicative approach) or the exclusion 
(additive approach) of the field independent wavefunction renormalization $z_k$. In the multiplicative approach, the rescaled 
regulator $\hat R_k$ contains the rescaled wavefunction renormalization $\hat z_k$. In the additive approach, the frequency 
can be absorbed by the overall multiplicative constant of the rescaled regulator or can be chosen arbitrarily since it is a 
scale-independent free parameter of the model. Important to note, that the additive approach requires the use of the 
power-law regulator function. 

The phase structure should be independent whether we use the multiplicative or additive approaches for the definition of the 
regulator and of its parameters such as $b$. For example, one can choose $b=2$, see Ref. \cite{optimal_sg}.
Regarding the regulator-dependence we note that the linearised RG flow equations can always be obtained analytically for the 
power-law type regulator, thus, there is no real need for the use of the Litim-type one. The full RG flow (with higher harmonics) 
requires a numerical treatment anyway (even for the Litim-type regulator) and produces us a complete picture of the phase diagram.

Let us first discuss the linearized RG flow equations obtained by the additive approach of the regulator function where
dimensionless couplings ${\tilde u}_k = k^{-d} u_k$, and ${\tilde z}_k = k^{2-d} \hat z_k$ are introduced. In this case, the RG flow
equations have the following forms for $d=1$ dimension
\bea
\label{lin_d1_u}
(1+k \partial_k) {\tilde u}_k &=& 
\frac{1}{4 b \sin(\frac{\pi}{2b})} 
{\tilde z}^{\frac{1-2b}{2b}}_k  {\tilde u}_k
+ \ord{{\tilde u}^2_k} \\
\label{lin_d1_z}
(-1+k \partial_k) {\tilde z}_k &=&
- \frac{c_1(b)}{4\pi}  {\tilde z}^{\frac{5-4b}{2b}}_k {\tilde u}^2_k
+ \ord{{\tilde u}^3_k}
\eea
with $c_1(b) = \frac{5\pi(55+4b(5b-18))}{48 b^2 \sin[5\pi/(2b)]}$. Eqs.~\eq{lin_d1_u}, \eq{lin_d1_z} have a non-trivial fixed point at 
${\tilde z}_{\star} = [4 b \sin(\pi/(2b)]^{2b/(1-2b)}$, ${\tilde u}_{\star}^2 = (4\pi/c_1) {\tilde z}_{\star}^{3-5/(2b)}$. 
For $d=2$, linearized RG equations are
\bea
\label{lin_d2_u}
(2+k \partial_k) {\tilde u}_k &=& 
\frac{1}{4 \pi {\tilde z}_k} {\tilde u}_k
+ \ord{{\tilde u}^2_k} \\
\label{lin_d2_z}
k \partial_k {\tilde z}_k &=&
- \frac{c_2(b)}{8\pi}  {\tilde z}^{\frac{2-2b}{b}}_k {\tilde u}^2_k
+ \ord{{\tilde u}^3_k}
\eea
with $c_2(b) = \frac{2\pi(b-2)(b-1)}{3 b^2 \sin[2\pi/b]}$ and result in a KTB type (i.e. infinite order) phase transition with 
${\tilde z}_{\star} =1/(8\pi)$ \cite{ncut1,d2_sg}. Important to note that $c_2(b) > 0$ for $b>1$.
Finally, for $d=3$, linearized RG equations are
\bea
\label{lin_d3_u}
(3+k \partial_k) {\tilde u}_k &=& 
\frac{1}{8 \pi b \sin(\frac{\pi}{2b})}   {\tilde z}^{\frac{-1-2b}{2b}}_k {\tilde u}_k
+ \ord{{\tilde u}^2_k} \\
\label{lin_d3_z}
(1+ k \partial_k) {\tilde z}_k &=&
- \frac{c_3(b)}{8\pi^2}  {\tilde z}^{\frac{3-4b}{2b}}_k {\tilde u}^2_k
+ \ord{{\tilde u}^3_k}
\eea
with $c_3(b) = \frac{\pi(3+4b(b-2))}{16 b^2 \sin[3\pi/(2b)]}$. Due to the tree-level scaling of $\tilde z_k$, the non-trivial fixed point appears 
for $d<2$ and disappears for $d>2$ in the RG flow. However, a "turning point" can be identified for $d>2$ where the irrelevant coupling 
$\tilde u_k$ turns to a relevant one. For $d=3$ the turning point is at ${\tilde z}_{\star} = [24 \pi b \sin(\pi/(2b)]^{-2b/(1+2b)}$.

In the multiplicative approach where the definition of the regulator $R_k$ includes $z_k$ beyond LPA, the linearized RG flow 
equations obtained for the dimensionless couplings are almost identical to those obtained by the additive case. The prefactors of
the r.h.s of the flow equations and the power of the Fourier amplitude $\tilde u$ are identical in the multiplicative and additive cases.
The difference is due to the power of the dimensionless wavefunction renormalization $\tilde z_k$. In the multiplicative case,
the r.h.s of the flow equations of the Fourier amplitude contain $\tilde z^{-1}_k$ and the flow equations of the wavefunction 
renormalization have $\tilde z^{-2}_k$ independently of choice of the dimension $d$ and the parameter $b$. Since the flow 
equations of the multiplicative and additive approaches should give the same phase structure, in this work we focus on the more
complex additive case.

\begin{figure}
\begin{center}
\includegraphics[width=0.5\linewidth]{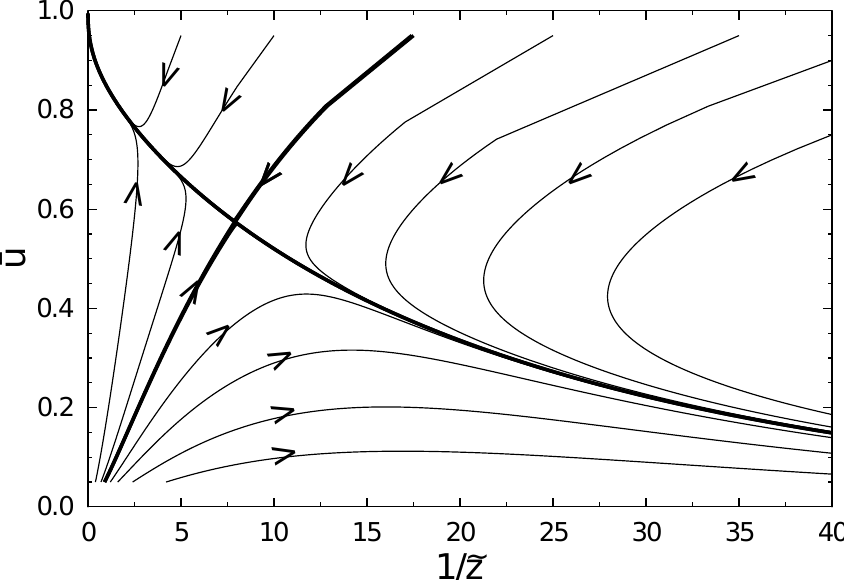}
\caption{
The exact phase diagram of the SG and the equivalent CG models for $d=1$ dimensions (similar RG flow can be drawn for $d<2$). Arrows 
indicate the direction of the flow. 
} 
\label{d1_flow}
\end{center}
\end{figure}

\begin{figure}[ht] 
\begin{center} 
\includegraphics[width=0.5\linewidth]{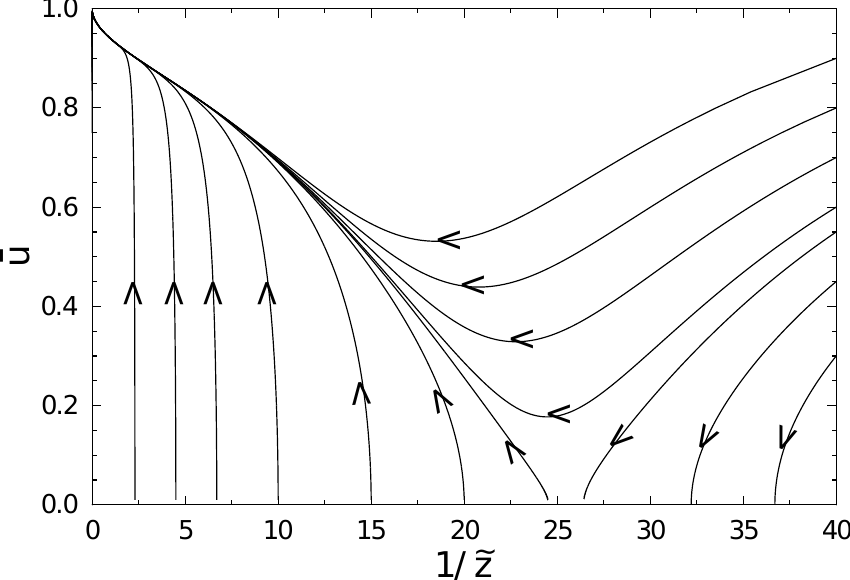}
\caption{
The exact phase diagram of the SG and the equivalent CG models for $d=2$ dimensions which shows the Kosterlitz-Thouless-Berezinski type 
phase transition.
} 
\label{d2_flow}
\end{center}
\end{figure}

\begin{figure}
\begin{center} 
\includegraphics[width=0.5\linewidth]{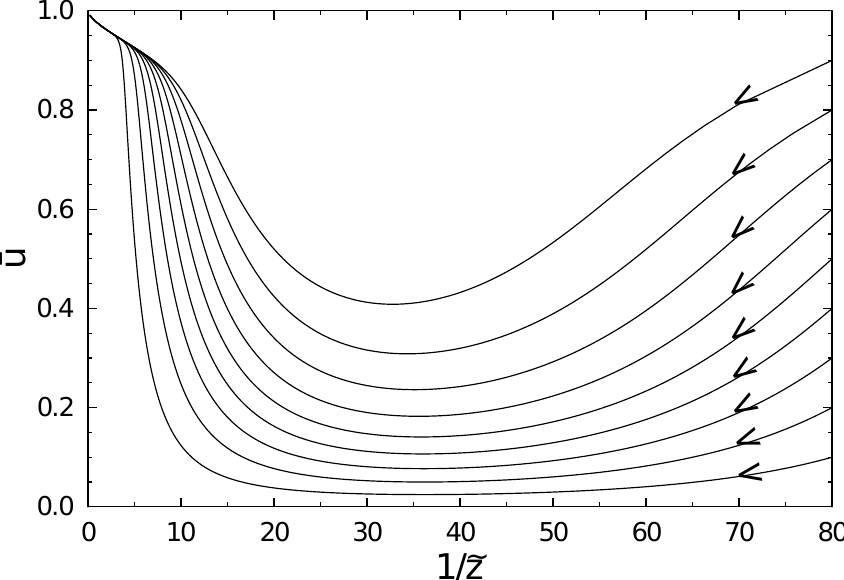}
\caption{
The exact phase diagram of the SG and the equivalent CG models for $d=3$ dimensions which indicates a single phase for $d>2$. 
} 
\label{d3_flow}
\end{center}
\end{figure}

In general, RG equations \eq{exact_u} and \eq{exact_z} (exact for a single Fourier mode) have to be solved numerically. Important 
feature of the exact RG flow the emergence of a new low-energy/infrared (IR) fixed point related to the degeneracy of the blocked 
action. Namely, Eqs. \eq{exact_u}, \eq{exact_z} become singular at the momentum scale where $\bar{k} - k^{2-d} u_k = 0$ 
with $\bar{k} = \min_{p^2} P = b k^2 [{\tilde z}_k/(b-1)]^{1-1/b}$. Therefore, it is convenient to redefine the dimensionless coupling 
constant as $\bar{u}_k \equiv  k^{2-d} u_k/\bar{k} = k^2{\tilde u}/\bar{k}$ which tends to one in case of degeneracy. Exact RG equations 
\eq{exact_u}, \eq{exact_z} were solved for $b=2$ and flow diagrams are plotted in \fig{d1_flow} for $d=1$, in \fig{d2_flow} for $d=2$, 
and in \fig{d3_flow} for $d=3$ dimensions. 
%
%
%
%
%
%
The IR fixed point ($\bar{u}_\star = 1$, $1/\tilde z_\star = 0$) which corresponds to degeneracy was found in any dimensions. The $d=1$ case 
needs further improvement because in quantum mechanics spontaneous symmetry breaking is not allowed. The appearance of the broken 
phase, was found for the anharmonic oscillator, see e.g. \cite{d1_anharmonic} where the higher order terms in the gradient expansion handle 
the problem.
 
Let us emphasize that only the exact RG flow with the inclusion of the wave function renormalization is suitable for the determination
of the IR fixed point since the perturbative (truncated) RG equations are non-singular \cite{trunc_rg_sg}. Thus any previous attempts based 
on perturbation theory \cite{trunc_rg_sg} or by the usage of the local potential approximation \cite{d3_sg,d3_sg_direct,d4_sg} were unable 
to determine the IR behavior in a reliable manner. Furthermore, the exact RG flow changes the position of the non-trivial fixed point obtained 
for $d=1$. Indeed, for $b=2$, linearized RG equations \eq{lin_d1_u}, \eq{lin_d1_z} result in $1/{\tilde z}_{\star} = 2^{10/3}\approx 10.08$ and 
$\bar{u}_{\star} = {\tilde u}_{\star}/(2 \sqrt{{\tilde z}_{\star}}) \approx 0.23$ which have been modified by the exact RG, 
see \fig{d1_flow}.  Similarly, the turning point of the $d=3$ case determined by the exact RG coincides with that of the linearized RG which is 
$1/{\tilde z}_{\star} = 4 (6\pi)^{4/5} \approx 41.9$ only for small $\bar{u}$, see \fig{d3_flow}. For $d=2$ dimensions the critical value 
$1/\tilde z_c = \beta_c^2 = 8\pi$ which separates the phases of the model (see \fig{d2_flow}) is found to be scheme-independent. 
The comparison of exact and perturbative (linearized) RG flow can be used for optimizing the renormalization schemes. 

Finally, let us discuss the possibility of having a topological phase transitions in higher dimensions. In \cite{extended_d4_sg} 
it was argued that the extended version of the SG model posseses a topological phase transition in 
$d=4$ dimensions. The kinetic terms of the extended model and the usual one \eq{sg} are different and this is 
the reason why one can observe a phase transition for the extended SG theory. Indeed, the $\Delta^{2}$ 
kinetic term is sub-leading with respect to the standard $\nabla^{2}$ term and, so, the topological (or KTB) phase 
transition in $d=4$ dimension is expected to be of "higher order" with respect to the standard one in $d=2$, 
and so, it needs a specific tuning of the coupling in the $\nabla^{2}$ in order to be observed. If one rescales the field 
by the frequency ($\varphi \to \beta \varphi$) in both versions of the SG model, the dimension of the wavefunction 
renormalization can be determined. In the extended SG model it is {\em dimensionless} which requires 
no tree-level scaling and it makes possible (under some other conditions) to observe a topological phase transition. 
However, the kinetic term of the usual SG model \eq{sg} has a tree-level scaling which has serious consequence, the 
RG flow has a turning instead of a critical point. The exact RG flow, see \fig{d4} in the following sections, confirms this 
expectation. So, if the usual kinetic term is generated by the RG flow, the absence of the topological phase transition is 
unavoidable in higher dimensions $d>2$.

\section{Double-frequency SG model}
\label{sec_dsg}
\begin{figure}
\begin{center} 
\includegraphics[width=0.5\linewidth]{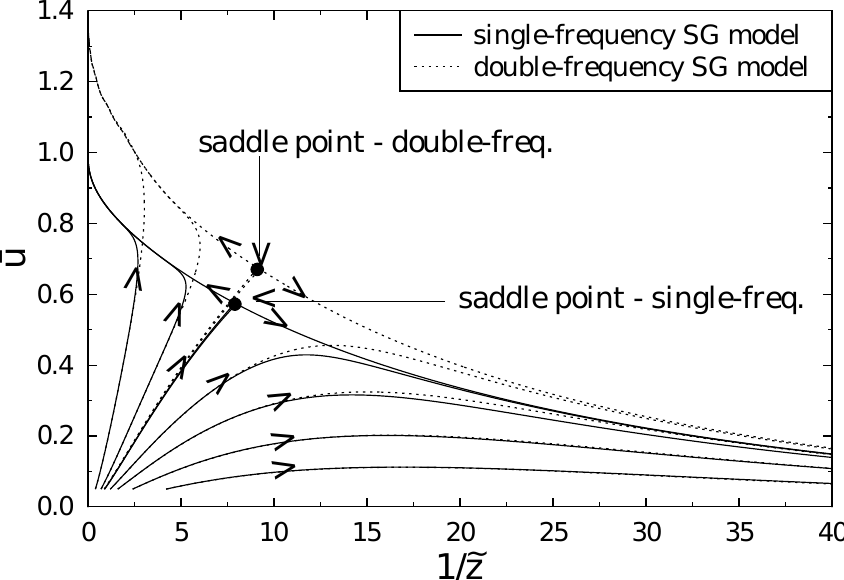}
\caption{
Lower half of the phase portrait of the SG model with single \eq{eaans_dimless} and double frequency \eq{dsg} are compared 
(in the $\bar u$, $1/\tilde z$ plane) for $d=1$.  Results are in qualitative agreement.
} 
\label{d1_double}
\end{center}
\end{figure}

Let us study the influence of higher harmonics on the phase structure of the single-frequency SG model. For $d=2$ the 
higher frequency modes of the SG model correspond to vortices with higher vorticity of the related XY model 
and it is known to play no role in the vortex dynamics  (the excitation of vortices with higher vorticity has lower probability). 
The equivalence between the SG and the XY models has been proven in $d=2$ dimensions, so it is important to clarify the 
results of the previous section which were obtained for the single-frequency SG model for $d\neq 2$. Let us consider the 
double-frequency SG model 
\bea
\label{dsg}
\Gamma_k = 
\int_x  \left[\frac{\bar z_k}{2} (\partial_\mu{\theta}_x)^2 
+ u_{1k} \cos(\theta_x)  +  u_{2k} \cos(2\theta_x)   \right]
\eea
where the first Fourier amplitude is equivalent to that of the single-frequency SG model, $u_{1k} \equiv u_k$. In case of the double-frequency model the
parameter space is three-dimensional thus its projection to the $\bar u, 1/\tilde z$ plane is used to compare RG trajectories of the 1-dimensional SG model 
with single and double frequency, see \fig{d1_double}. 
%
%
Let us note, the same normalized Fourier amplitude $\bar u$ is used both for the single- and the double-frequency cases. The definition of 
$\bar u$ is given by the low-energy behavior of the inverse propagator (see the previous section) and it is normalized in such a way that for the 
single-frequency model it tends to 1.0 in the low-energy limit of the broken phase. The low-energy limit of $\bar u$ in the broken phase of 
the double-frequency SG model is similar to that of the single-frequency model. Similarly, a non-trivial saddle point appears in $d=1$ dimension 
in the RG flow of the SG model both for the single- and double-frequency cases but its position differs from each other, see \fig{d1_double}. Let us 
also mention that the numerical results shown in \fig{d1_double} are obtained by solving the RG equations \eq{ea_v} and \eq{ea_z} both for 
the single- and double-frequency SG models in $d=1$ using the same initial condition for $u_{1\Lambda} \equiv  u_\Lambda$ and for the 
double-frequency case the other Fourier amplitude is chosen as $u_{2\Lambda} = 0.1 u_{1\Lambda}$.

The main conclusion is that results of the RG analysis of the single- and double-frequency SG models are in qualitative agreement which 
demonstrates that similarly to the $d=2$ case, the higher harmonics do not modify the phase structure given by the single-frequency model.

\section{Phase structure of the neutral CG}
\label{sec_cg}
The CG model has been the subject of an intense study in last decades \cite{d_cg_general,d_cg,sg_cg,trunc_rg_sg,d3_sg,samuel} 
and there is a continuous interest in the use of the SG representation of CG systems \cite{sg_cg, trunc_rg_sg,d3_sg,samuel} 
since it has an important relevance in various models of critical phenomena (e.g. superfluid transition in He films, melting of crystals in $d=2$, 
arrays of Josephson junctions, roughening transition). However, the complete understanding of the critical behaviour of the d-dimensional 
neutral CG which is a plasma of equal number of positive and negative charges is still lacking. For example, previous investigations 
(even for the well studied 2-dimensional CG model) were restricted to the case of low densities (limit of low fugacity) and there are different 
competing views of what happens to the neutral plasma at high densities \cite{d_cg_general,sg_cg}. Thus, an RG study of the d-dimensional 
CG beyond the dilute gas approximation is required which is one of the goal of the present work. 

Since the mapping between the CG and SG models holds in arbitrary dimension \cite{samuel, d3_sg} (and it is exact in case of point-like charges) 
the RG study of the d-dimensional SG model can be directly used to map out the phase structure of the CG model. In the framework of perturbation 
theory (limit of low fugacity) RG equations of the CG model in arbitrary dimension is given in \cite{d_cg} and reads as
\bea
\label{cg_rg}
\frac{dx}{dl} = -x(x y^2 + d - 2), \,\,\,
\frac{dy}{dl} = -y(x-d) 
\eea
with $\partial_l = -k \partial_k$, $y \sim {\tilde u}_k$ and $x \sim 1/{\tilde T} \sim 1/{\tilde z}_k$ where ${\tilde T}$ is the temperature and $y$ is the 
fugacity. Using these identities Eq. \eq{cg_rg} can be rewritten as
\begin{equation}
\label{cg_uz}
((d-2) + k\partial_k) {\tilde z}_k = - c_z {\tilde u}^2_k, \,\,\,
(d + k\partial_k) {\tilde u}_k =  c_u \frac{1}{{\tilde z}_k} {\tilde u}_k,
\end{equation}
with constants $c_u$, $c_z$ and they are found to be similar to the linearized RG equations obtained for the SG model, 
see Eqs.~\eq{lin_d1_u},\eq{lin_d1_z}, Eqs.~\eq{lin_d2_u},\eq{lin_d2_z} and Eqs.~\eq{lin_d3_u},\eq{lin_d2_z}. For example,
the non-trivial fixed point of the 1-dimensional SG model can be identified in the flow generated by \eq{cg_uz}, too. In the 
so called sharp cutoff limit ($b\to \infty$) the linearized RG equation derived for the Fourier amplitude of the SG model reduces 
to $(d + k\partial_k) {\tilde u}_k = \Omega_d {\tilde u}_{k}/({\tilde z}_{k}2(2\pi)^d)$ which is identical to the second equation of \eq{cg_uz}. 
However, the sharp cutoff does not support to the derivative expansion at higher order, i.e. for $b\to \infty$ the multiplicative constant 
$c_d(b)$ of the RG equation obtained for ${\tilde z_k}$ becomes infinity. 

Most important new results of the exact RG flow are the existence of the high temperature ($\bar{u}_{\star} = 1$, $1/{\tilde z}_{\star} =0$) and the 
absence of new further non-trivial fixed points. Let us note that, the perturbative (truncated) RG has a non-singular structure hence it is not able to 
recover the degenerate potential therefore it is not suitable to observe the high temperature fixed point \cite{d_cg}. Moreover, the exact RG changes 
the position of the non-trivial fixed point for $d<2$ and the turning point for $d>2$. Furthermore, since the mapping between the SG and CG models 
is exact for point-like charges the exact RG flow indicates a single phase for the CG for $d>2$.

\section{Checking the results}
\label{sec_check}
So far RG equations beyond the local potential approximation (LPA) have been derived for the SG model (with single and double frequency) in 
d-dimensions. By using the mapping of the SG theory onto the CG model, its RG flow and phase structure has been discussed beyond the 
dilute gas approximation. In this section let us check the findings of the present work by comparing them to known recent results 
\cite{barkhudarov,malard}.

Let us first rewrite the linearised RG flow equations \eq{lin_d1_u}, \eq{lin_d1_z}, \eq{lin_d2_u}, \eq{lin_d2_z} and \eq{lin_d3_u}, \eq{lin_d3_z} in a 
single form valid in arbitrary dimensions
\bea
(d + k \partial_k) \tilde u_k = A_{b,d} \,\, \tilde z_k^{\frac{(2-d)-2b}{2b}} \, \tilde u_k  + \ord{{\tilde u}^2_k} \\
((d-2) + k \partial_k) \tilde z_k = - B_{b,d} \,\, \tilde z_k^{\frac{(6-d)-4b}{2b}} \, \tilde u_k^2  + \ord{{\tilde u}^3_k}
\eea
where $A_{b,d}$ and $B_{b,d}$ are constants depend on the dimension $d$ and the regulator parameter $b$. For the comparison 
it is convenient to use the frequency instead of the wave function renormalization ($\tilde \beta^2 = 1/\tilde z$),
\bea
\label{d_lin_u_b}
k\partial_k \tilde u_k &=& 
\left[ A_{b,d} \, (\tilde \beta_k^2)^{\frac{2b-(2-d)}{2b}} -d\right] \tilde u  \\
\label{d_lin_z_b}
k\partial_k \tilde \beta_k^2 &=& 
\tilde \beta_k^2 \left[(d-2)  + B_{b,d} \, \tilde u_k^2 (\tilde \beta_k^2)^{\frac{6b- (6-d)}{2b}} \right],
\eea
and take the limit $b\to \infty$ which is identical to the so called sharp cutoff regulator (as mentioned in the previous section), 
\bea
\label{d_lin_u}
k\partial_k \tilde u_k &=& 
\left[ A_{\infty,b} \, \tilde \beta_k^2 -d\right] \tilde u  \\
\label{d_lin_z}
k\partial_k \tilde \beta_k^2 &=& 
\tilde \beta_k^2 \left[(d-2)  + B_{\infty,b} \, \tilde u_k^2 \tilde \beta_k^6 \right].
\eea
It is important to note that $B_{\infty,b}$ cannot be defined unambigously in the functional RG approach since 
the sharp cutoff confront to the derivative expansion, however, it is possible in the real space RG even beyond LPA. 

Let us first compare Eqs. \eq{d_lin_u} and \eq{d_lin_z} to the RG equations (3.2.8) and (3.2.9) of \cite{barkhudarov} which is obtained in the 
dilute gas approximation and reads
\bea
\label{barkhudarov_u}
k\partial_k \tilde u_k &=& 
\left[ \frac{K_d}{2} \, \tilde \beta_k^2 -d\right] \tilde u  \\
\label{barkhudarov_z}
k\partial_k \tilde \beta_k^2 &=& 
\tilde \beta_k^2 \left[(d-2)  + \left(\frac{I_1 K_d}{2} +B\right) \, \frac{\tilde u_k^2}{4} \tilde \beta_k^6 \right].
\eea
by using the following identifications $2z \equiv \tilde u$, $\alpha \equiv \beta$ and $\partial_l \equiv -k\partial_k$ where $K_d$, $I_1$ and $B$ 
are constants. It is clearly demonstrated that Eqs. \eq{d_lin_u} and \eq{d_lin_z} are identical to \eq{barkhudarov_u} and \eq{barkhudarov_z} up 
to some constant.

Finally, let us consider the flow equations above (45) in \cite{malard} which are obtained for the SG model in $d=2$ using the Wilson-Kadanoff 
blocking relation up to leading order terms and reads as
\bea
\label{malard_u}
k\partial_k \tilde u_k &=& 
\left(\frac{\tilde \beta_k^2}{4\pi} -2\right) \tilde u  \\
\label{malard_z}
k\partial_k \tilde \beta_k^2 &=& 
\frac{3 \tilde \beta_k^6 \tilde u_k^2}{4 \pi \Lambda^3}
\eea
where the identifications $g \equiv \tilde u$ and $\partial_l \equiv -k\partial_k$ are used. One finds agreement between 
Eqs. \eq{d_lin_u_b} \eq{d_lin_z_b} and Eqs. \eq{malard_u} \eq{malard_z} since $A_{b,2} = 1/(4\pi)$ and $b=2$ is chosen.
Thus, it was shown that the findings of the present work recovers known (approximate) results which validates our conclusions.

\section{Isotropic classical XY spin model}
\label{sec_xy}
The partition function of the d-dimensional isotropic classical XY spin model \cite{zj} can be expressed in terms of topological excitations of the 
original degrees of freedom. For $d=2$ dimensions, the dual theory is the vortex gas which is known to belong to the class of universality of the neutral 
CG \cite{nienhuis,huang,lattice_cg}. Two-dimensional generalized models are well known where both the CG and the vortex gas are included as particular 
limiting cases \cite{nienhuis,huang,lattice_cg} and are self-dual under the duality transformation. For $d=3$, the dual theory is the gas of interacting 
vortex loops \cite{nelson,shenoy} (i.e. the lattice CG \cite{lattice_cg}). Corresponding flow equations have been derived for the parameters $K$ 
(i.e. the coupling between the spins) and $y$ (i.e. the fugacity of the vortex loops) by real-space RG method ($a \sim 1/k$ where
$a$ is the running cutoff in the coordinate space while $k$ is the running momentum cutoff) in the limit of low fugacity \cite{d3_sg,nelson,shenoy}, 
\begin{eqnarray}
\label{xy_rg}
a\partial_a K = K-  \frac{4\pi^3}3 K^2 y,~~~~
a\partial_a y = \left( 6- \pi^2 K L \right) y,
\end{eqnarray}
where $L$ approaches a constant in the IR limit ($L\to 1$) \cite{nelson}, or it is weakly divergent ($L \to  \ln(a/a_c ) + 1$) \cite{shenoy}. 
Since $a\partial_a = - k\partial_k$ and by using the identities $y \equiv \tilde u$, $K\equiv 1/{\tilde z}$, RG equations of \eq{xy_rg} are rewritten as
\begin{eqnarray}
\label{xy_uz}
k\partial_k {\tilde z}_k = {\tilde z}_k -  \frac{4\pi^3}3 {\tilde u}_k^2, \,\,\,\,
k\partial_k {\tilde u}_k = 
\left(\frac{1}{{\tilde z}_k} \frac{\pi^2 L}{2} - 3 \right) {\tilde u}_k,
\end{eqnarray}
which can be compared to the linearized RG obtained for the SG \eq{lin_d3_u}, \eq{lin_d3_z} and the CG \eq{cg_uz} models in $d=3$ dimensions. 
The only qualitative disagreement is the sign of the tree-level scaling term of ${\tilde z}_k$ but it has important consequences 
since the RG flow of the vortex-loop gas has a non-trivial fixed point which is absent in case of the 3-dimensional SG and CG models. 
Therefore, it demonstrates that the vortex-loop gas has a different scaling behaviour, thus it belongs to a class of universality 
different from that of the SG and CG models for $d\neq 2$. For $d=2$, the couplings $K, {\tilde z}$ have no tree-level scaling 
thus SG, CG, XY models are in the same universality class. Let us emphasize that RG equations are compared at the linearized level, 
however the exact RG study of the SG and CG models is required since it shows the absence of new non-trivial fixed points, thus there is 
no room to find a mapping between the parameters of the vortex-loop gas and the 3-dimensional CG which could produce the same phase structure.

\section{Application to Higgs, inflaton and axion physics}
\label{sec_appl}
\begin{figure}
\begin{center} 
\includegraphics[width=0.5\linewidth]{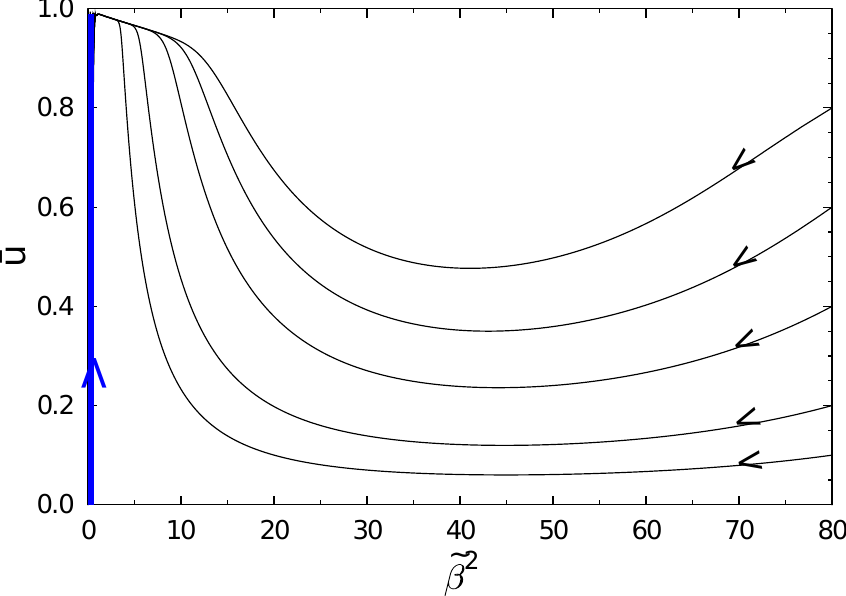}
\caption{
Phase diagram of the SG model in $d=4$ dimensions.} 
\label{d4}
\end{center}
\end{figure}

In this section we apply the results of the functional RG study of the SG model obtained in $d=4$ dimensions for Higgs, inflaton and axion physics.
The phase diagram of the four-dimensional SG model is shown in \fig{d4}. 
%
%

Let us first discuss its consequences on the stability of the periodic Higgs potential in the framework of the Higgs-Yukawa model 
\cite{higgs_frg}
\beq
\Gamma_k = \int_k \left[\hf Z_{\phi,k} (\partial_\mu \phi)^2 + U_k(\rho) + Z_{\psi,k} \bar\psi i \slashed{\partial} \psi  + i h_k \phi \bar\psi \psi \right]
\eeq
where the Yukawa coupling ($h_k$) has been introduced in order to couple the Higgs field ($\phi$) to a fermion field 
($\psi$) (top quark). The scalar potential is assumed to be a periodic one, i.e., $U_k(\rho) \sim \cos(\beta\rho)$ with $\rho=\hf \phi^2$. 
The functional RG equation suitable for the Higgs-Yukawa model reads as \cite{higgs_frg}
\beq
k\partial_k\Gamma_k = \hf\mr{STr} \frac{k\partial_k R_k}{R_k+\Gamma_k^{(2)}} \, .
\eeq
If the Yukawa coupling contains a linear term in the Higgs field (which is the usual case) and the scalar potential periodic in the field 
components, then the Hessian of the running action $\Gamma_k^{(2)}$ with respect to the fields ($\phi, \psi, \bar\psi$) has a fully periodic 
part. In other words, the free boson contribution for the functional RG equation remains periodic and the application of the RG study of the 
pure periodic scalar model on the stability problem \cite{NNLO_stability,bezrukov_rubio_shapo} represents a reliable approximation. 
Although, the scalar potential is periodic in the magnitude, i.e., in $\rho$ which requires a more careful analysis  where one has to treat the 
RG running of every couplings in a complete study but this is not the goal here. Coming back to the RG results of the pure SG model 
the following statements can be done. Since the dimensionless Fourier amplitude ($\tilde u_k$) tends to a constant value (see, \fig{d4}) 
the dimensionless periodic potential remains bounded from below and above which guarantee the stability at this level of approximation. 
A polynomial potential can loose stability if it becomes unbounded from below, but a periodic potential with finite Fourier amplitude
cannot be unstable. This is why we argue that stability of the periodic scalar potential is guaranteed because the RG study of the SG model 
results in a finite value for the amplitude. Furthermore, it also means that the dimensionful Fourier amplitude (and consequently the 
dimensionful mass term for the Higgs) tends to a small number in the deep IR limit. This is required since the Higgs mass 
at the scale of inflation (UV scale) should be large fixed by Cosmic Microwave Background Radiation (CMBR) data and 
it should tend to the measured Higgs mass in the low-energy, IR limit which orders of magnitude smaller than at the UV scale. Thus, the 
RG result of the pure periodic model supports to consider the periodic model as a possible UV completion of the SM Higgs potential. 
 
\begin{figure}
\begin{center} 
\includegraphics[width=0.5\linewidth]{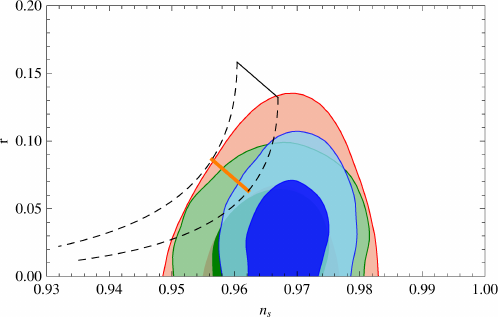}
\caption{
Orange line represents the best fit of the periodic inflationary potential (for $N\in[50,60]$) with respect to its frequency to Planck 
results \cite{planck}. This gives the optimal value which is $\beta = 0.15$. The black line stands for the quadratic inflationary 
potential (for $N\in[50,60]$) which is almost excluded. Dashed lines indicate allowed regions for the periodic model (for various 
frequencies where $\beta = 0$ recovers the quadratic potential).} 
\label{sg_planck}
\end{center}
\end{figure}

Let us turn to the discussion of the RG running of the (dimensionless) frequency $\tilde \beta$ of the periodic inflationary potential, 
for the details see \app{sg_inflaton}. Conditions for the slow-roll-down inflation requires a small frequency ($\tilde \beta = 0.15$) 
which is shown by \fig{sg_planck}.
%
%
The blue line of \fig{d4} stands for the RG trajectory for such a small frequency. It is an almost vertical straight line, i.e. $\tilde \beta$ 
does not change over inflation and in the post-inflationary period. Moreover, it tends to zero in the deep IR limit which allows the 
Taylor expansion of the periodic potential and results in a simple quadratic potential. The RG study presented here, 
can be applied for a modified version of the periodic potential in which, in addition to the periodic self-interaction, a linear term appears
\beq
V = g \phi + u \cos(\beta\phi) 
\label{lin_plus_sg}
\eeq
which was considered in \cite{linear_periodic} as a new type of inflationary potential. The idea is similar to the Higgs case, the Tr-term 
of the functional RG equation remains periodic since not the potential but its Hessian appears there which is periodic even for the 
above potential. 

Finally, let us consider the consequences of the RG study of the SG model on axion physics. In Ref. \cite{axion_flattening} it was
shown that RG transformations results in a flat (dimensionful) axion potential at LPA. Findings of this work are suitable to go beyond LPA. 
Indeed, the flow diagram \fig{d4} clearly takes into account the (non-trivial) RG running of the frequency which can be seen only beyond LPA.
Since the dimensionless Fourier amplitude tends to a constant value the corresponding dimensionful parameter vanishes in the IR
limit, the dimensionful potential flattens out. This confirms the flattening of the axion potential beyond LPA.

\section{Summary}
\label{sec_sum}
In this work the functional RG study of the periodic scalar field theory, i.e. the sine-Gordon (SG) model has been performed in 
arbitrary dimension and the wave function renormalization is included. This allows us to go beyond approximations used in previous 
works and to consider the consequences of the RG study of the periodic scalar model on Higgs, inflaton and axion physics
and to consider the possibility of a topological phase transition in higher dimensions.

For example a periodic Higgs potential was proposed as a possible extension of the standard quartic one. Using the usual 
parametrisation of the field around the first minimum, one recovers the Lagrangian of the Higgs sector but with a periodic type 
self-interaction term for a single-component real scalar field. The exact phase structure of the pure SG scalar field theory (with no 
gauge and fermionic fields) was mapped out by the functional RG method in arbitrary dimension. It was found that the SG model 
has a single phase in $d>2$ and the potential remain bounded from below. This result opens a new platform to perform stability 
studies for models with periodic type Higgs potentials, as an example one can mention the so called massive sine-Gordon model
which contains quadratic and periodic terms \cite{pseudo_periodic_higgs_inflation}.

The RG running of the frequency parameter is used to answer to the question whether the conditions for slow-roll inflation 
remains valid under RG transformations. It was shown that starting from a small value for the frequency (which is required), it 
does not evolve along the corresponding RG trajectory. Thus, our analysis supports the viability of a periodic inflationary potential.

Finally, it was also shown that the dimensionful periodic potential flattens out in the IR limit which confirms previous results on the 
flattening of the axion potential. 

The findings were used to determine the phase structure of the equivalent neutral Coulomb-gas (CG) too. The high temperature fixed point of 
the d-dimensional SG and CG was determined and shown that the perturbative RG is not suitable to recover it. The position of the non-trivial 
fixed point of the RG flow for $d<2$ and the turning point for $d>2$ dimensions given by the exact flow is compared to the linearized one. 
Note, that for $d=1$ dimension, the broken phase should be vanish if the approximations used are improved \cite{opt_d1sg}.

The absence of new further non-trivial fixed points were also demonstrated which results in a single phase for the SG scalar theory and the 
neutral CG with point-like charges for $d>2$, thus, (i) no further phase transition can be identified at high densities for the neutral CG as 
opposed to e.g. \cite{d_cg_general,sg_cg}, (ii) no topological phase transition can be observed for the SG model with the usual
kinetic term \eq{sg} in higher dimension $d>2$, (iii) there is no way to find a mapping between the phase structure of the neutral 
3-dimensional CG and the vortex-loop gas which is the gas of topological defects of the isotropic XY spin model for $d=3$, i.e., 
they belong to different universality classes.

\section*{Acknowledgements}

This work was supported by the J\'anos Bolyai Research Scholarship of the Hungarian Academy of Sciences. Useful discussions with
N. Defenu, D. Horvath, G. Somogyi, Z. Trocsanyi, A. Trombettoni are gratefully acknowledged.

\appendix

\section{Periodic inflationary potential}
\label{sg_inflaton}
A possible application of the higher dimensional periodic, i.e. sine-Gordon type scalar field theory is inflationary cosmology.
Let us start with the Friedman-Robertson-Walker (FRW) metric $g_{\mu\nu} = \mr{diag}(-1,a^2,a^2,a^2)$ using natural units
$c = \hbar = 1$ where $a=a(t)$ is the scale factor of which time-dependence is determined by the Friedman equation (derived form the
Einstein equation for the FRW metric). The solution of the Friedman equation with the assumption for the equation of state $p= -\rho$
results in exponential expansion, i.e., $a(t) \sim e^{Ht}$ where $H$ is the Hubble constant. The key observation is that scalar fields 
can mimic the required equation of state (under some conditions) thus represent excellent models for inflation \cite{encyc}. The 
Lagrangian of the Einstein-Hilbert action involving a real scalar field minimally coupled to gravity is given by 
\beq
S=\int d^4x \sqrt{-g} \left[\frac{m_p^2}{2} R +{\cal L}_\phi \right], \hskip 0.2cm
{\cal L}_\phi=-\hf g^{\mu\nu} \partial_\mu \phi \partial_\nu \phi -V(\phi) = \hf \dot \phi^2-\hf \frac{1}{a^2} (\nabla\phi)^2 -V(\phi)
\nonumber
\eeq
with $\sqrt{-g}=\sqrt{-\det(g_{\mu\nu})}=a^3$. The first observation is that over inflation the field can be considered to be 
homogeneous ($\nabla \phi/a = 0$). The simplest example for inflationary potential is the so-called chaotic monomial or large
field inflation where the potential reads
\beq
V=\lambda m_p^{4-\alpha} \phi^\alpha, \hskip 1cm
m_p^2=\frac{1}{8\pi G}, \hskip 1cm
m_p = 2.4 \times 10^{18} {\mr{GeV}}
\nonumber
\eeq
where $m_p$ is the (reduced) Planck mass, G is the gravitational constant and $\alpha=2,4$ is a typical choice. Conditions for the 
potential can be derived in order to fulfil the requirements of prolonged inflation (with slow roll down). These conditions are written 
in terms of parameters ($\epsilon$, $\eta$ and $N$) which are the following for chaotic or large field inflation (LFI)
\begin{align}
\epsilon=&\hf m_p^2 \left( \frac{V'}{V} \right)^2
= \hf m_p^2 \left( \frac{ \lambda \alpha m_p^{4-\alpha} \phi^{\alpha-1}}{\lambda m_p^{4-\alpha} \phi^\alpha} \right)^2
=\frac{m_p^2}{2} \frac{\alpha^2}{\phi^2} \ll 1
\nonumber
\\
\eta=& m_p^2 \frac{V''}{V}
=m_p^2 \left( \frac{\lambda \alpha (\alpha-1) m_p^{4-\alpha} \phi^{\alpha-2} }{\lambda m_p^{4-\alpha} \phi^\alpha} \right)
=m_p^2 \frac{\alpha(\alpha-1)}{\phi^2} \ll 1
\nonumber
\\
\label{Nmonomial}
N=&-\frac{1}{m_p^2} \int_{\phi_i}^{\phi_f} d\phi \frac{V}{V'}
=-\frac{1}{m_p^2} \int_{\phi_i}^{\phi_f} d\phi \frac{\phi}{\alpha}
=\frac{1}{2 \alpha m_p^2} (\phi_i^2-\phi_f^2) = [50 - 60]
\nonumber
\end{align}
Since inflation ends when $\epsilon\approx1$ or $\eta\approx1$ therefore one has to choose 
\beq
\phi_f=\alpha m_p  \implies \epsilon=\hf \hskip0.5cm \eta=\frac{\alpha-1}{\alpha}, 
\hskip 1cm
\phi_i=\sqrt{2\alpha m_p^2 N +\phi_f^2}\approx \sqrt{2\alpha N} m_p 
\nonumber
\eeq
where at the last approximation we used that $\phi_i \gg \phi_f$. From the above parameters one can build up measurable quantities 
such as the spectral index ($n_s \approx 2\eta - 6\epsilon$) and the tensor-to-scalar ratio ($r \approx 16 \epsilon$) which relates to 
each other. For the quadratic case ($\alpha = 2$) the relation  is independent of $N$ and valid up to order $(1/N)$ 
\beq
\boxed{(n_s -1) + \frac{r}{4} = 0},
\nonumber
\eeq
and can be compared to observations directly. For example, result of the Planck mission \cite{planck} put strong constraints on the 
applicability of the quadratic LFI model, i.e., it is almost excluded by measured data on the thermal fluctuations of the cosmic 
microwave background radiation (CMBR). Indeed, testing of competing scenarios for scalar inflationary models against observation 
is a general issue, see for example \cite{inflaton_testing,inflaton_testing_2}.

Let us now turn to our original plan, i.e., to use a periodic scalar field as an inflationary model. Consider the sine-Gordon model in 
the following representation which is the so called Pseudo-Nambu-Goldstone boson potential (PNGB) \cite{periodic_inflaton}, 
\beq
V=u\left[1- \cos(\beta \phi)\right]
\nonumber
\eeq
where $u$, $\beta$ are dimensionful parameters. Generalization of that can be find also in Ref. \cite{linear_periodic}.
In $d=4$ dimensions the scalar field carries a dimension, $\phi = k^{(d-2)/2} \tilde \phi$ where $\tilde \phi$ is dimensionless and 
$k$ is an arbitrary chosen momentum scale convenient to take at the planck mass $k = m_p$. Thus, the corresponding 
dimensionless parameters are $\beta = m_p^{-1} \tilde \beta$ and $u = m_p^{4} \tilde u$. Calculating $\epsilon$, $\eta$ 
and $N$ parameters one finds
\begin{align}
\epsilon=&\hf  m_p^2 \left( \frac{V'}{V} \right)^2
= \hf \tilde \beta^2 \frac{\sin^2(\tilde \beta \tilde \phi)}{[1-\cos(\tilde \beta \tilde \phi)]^2} \ll 1
\nonumber
\\
\eta=&  m_p^2 \frac{V''}{V}
=  \tilde \beta^2 \frac{\cos(\tilde \beta \tilde \phi)}{1-\cos(\tilde \beta \tilde \phi)} \ll 1
\nonumber
\\
\label{Nmonomial}
N=&- \frac{1}{m_p^2} \int_{\phi_i}^{\phi_f} d\phi \frac{V}{V'}
= -\frac{2}{\tilde \beta^2}  \log \cos\left(\frac{\tilde \beta \tilde \phi}{2}\right) \Big|^{\tilde \phi_i}_{\tilde \phi_f} = [50-60]
\nonumber
\end{align}
and assuming a small value for the frequency $\tilde \beta$ the conditions can be fulfilled \cite{periodic_inflaton_test} 
which validates the use of the periodic scalar field theory in inflationary cosmology for small frequencies $\tilde \beta \ll 1$. 
In this limit one can expand the periodic potential and finds \cite{periodic_inflaton_test}
\beq
\boxed{(n_s -1) + \frac{r}{4}  = -\tilde \beta^2}
\nonumber
\eeq
where the last equation can be compared to the that of the quadratic potential. Precise measurements on $n_s$ and $r$ can 
be used to put constraints on the dimensionless frequency $\tilde \beta$ \cite{periodic_inflaton_test}. The important difference 
between the quadratic and the periodic cases is the presence of the frequency in the above formula which is  a running 
parameter of the model thus renormalization effects should be taken into account at least in the post-inflation period.

Indeed, one of our goal is to determine the RG running of the pure sine-Gordon theory in $d=4$ dimensions and to discuss the 
consequences on the use of the periodic model to inflationary physics.

\section{Periodic Higgs potential}
\label{periodic_higgs}

In the standard model (SM) of particle physics the Brout-Englert-Higgs (BEH) mechanism \cite{englert_brout,higgs} has been used 
to generate mass for the weak gauge bosons. Above a criticial temperature the electroweak symmetry is unbroken, all elementary 
particles are massless. Below the critical temperature the symmetry is spontaneously broken and the W and Z bosons acquire 
masses. Furthermore, fundamental fermions can also acquire mass as a result of their interaction with the Higgs field.
The SM Higgs field is an SU(2) complex scalar doublet with four real components 
\beq
\phi = \frac{1}{\sqrt{2}} \left(
\begin{array}{c}
\phi_1 +i \phi_2\\
\phi_3 +i \phi_4\\
\end{array}
\right). \nonumber
\eeq
The underlying symmetry of the electroweak sector is $SU(2)_L \times U(1)_Y$, thus, the Higgs Lagrangian reads as
\bea
&{\cal L} = (D_\mu \phi)^\star (D^\mu \phi) - V(\phi) - \frac{1}{2} \Tr \, ({F}_{\mu\nu} {F}^{\mu\nu}),  \nn
&\hskip -0.35cm V = \mu^2 \phi^\star \phi + \lambda (\phi^\star \phi)^2, \,
(D_\mu = \partial_\mu + i g {\bf T} \cdot {\bf W}_\mu +i g' y_j B_\mu)   \nonumber
\eea
and the vacuum expectation of the Higgs field is either at zero field for $\mu^2>0$ or at 
$\sqrt{\phi^\star \phi} = \sqrt{-\mu^2/(2\lambda)} = v/\sqrt{2}$ for $\mu^2<0$. After the spontaneous symmetry breaking 
of $SU(2)_L \times U(1)_Y$ into $U(1)_Q$ the field around the groundstate can be parametrized as
\beq
\phi(x) = \frac{1}{\sqrt{2}} \exp{\left(i\frac{{\bf T} \cdot { \boldsymbol{\xi}}(x)}{v}\right)} \left(
\begin{array}{c}
0\\
v +h(x)\\
\end{array}
\right) 
\nonumber
\eeq
with $v = 246$ GeV known from low-energy experiments and the unitary phase can be dropped by choosing an appropriate gauge. 
As a consequence of the BEH mechanism the photon remains massless but the weak gauge bosons acquire masses.
Three degrees of freedom of the Higgs field (out of the four) mix with weak gauge bosons while the remaining degree of freedom 
becomes the Higgs boson which is discovered at CERN's Large Hadron Collider. The complete Lagrangian for the Higgs sector 
reads
\beq
{\cal L} = \frac{\partial_\mu h \partial^\mu h}{2} - \frac{M^2_h h^2}{2} - \frac{M^2_h h^3}{2v}  - \frac{M^2_h h^4}{8v^2}  
+ \left(M^2_W W^+_\mu W^{-\,\mu} + \frac{M^2_Z Z_\mu Z^\mu}{2} \right) \left(1+\frac{2h}{v}+\frac{h^2}{v^2}\right)
\nonumber
\eeq
with $M_h = \sqrt{-2\mu^2} = \sqrt{2\lambda v^2}$. The measured value for the Higgs mass $M_h = 125.6 \, \mr{GeV}$ \cite{ATLAS,CMS}
implies $\lambda = 0.13$. This is close to the predicted value based on the assumption of the absence of new physics between the 
Fermi and Planck scales and the asymptotic safety of gravity \cite{shapo_wett}. This enables us to extrapolate the SM up to very high
energies and to interpret the Higgs boson as the inflaton by using two competing scenarios, (i) non-minimal coupling to gravity, 
(ii) higgs-inflation from false vacuum.

Let us first discuss the large non-minimal coupling to gravity \cite{higgs_inflation_1}, which results in the following action 
\beq
S = \int d^4x \sqrt{-\bar{g}} \frac{m_p^2}{2} \left[ (1+\xi \tilde h^2) \bar{R} -\bar{g}^{\mu\nu} \partial_\mu \tilde h \partial_\nu \tilde h - 2
m_p^2 \frac{\lambda}{4} \left(\tilde h^2 +  \frac{2v^2}{m^2_p}\right)^2 \right]
\nonumber
\eeq
where $\bar{g}^{\mu\nu}$ is the metric in the Jordan frame, $\xi$ represents the coupling between gravity and the dimensionless 
Higgs scalar $\tilde h$. To perform the slow-roll study, the action is usually rewritten in the Einstein frame where it takes the form 
\beq
S = \int d^4x \sqrt{-g} \left[m_p^2 \frac{R}{2} - \hf g^{\mu\nu} \partial_\mu \phi \partial_\nu \phi - 
\frac{m_p^4 \lambda}{4 \xi^2} \left(1 - e^{-\sqrt{2/3} \phi/m_p} \right)^2 \right] 
\nonumber
\eeq
in which the metric tensor being denoted by $g^{\mu\nu}$ and the new field variable is introduced via 
$d\phi/d\tilde h = m_p \sqrt{1 + \xi(1 + 6\xi)\tilde h^2}/(1+\xi \tilde h^2)$ but as a possible drawback of the method, 
perturbative unitarity can be violated for $\xi \neq 0$.

In case of a Higgs inflation from false vacuum a minimal coupling is used, i.e., $\xi = 0$ which implies $\phi = m_p \tilde h$
with the same shape of potentials in both frames but the SM Higgs potential is extended and assumed to develop a second 
(or more) minimum \cite{higgs_inflation_2,pseudo_periodic_higgs_inflation}. 
This scenario can have difficulties to achieve an exit from the inflationary phase. 
In addition, the measured Higgs mass is close to the lower limit, $126$ GeV, ensuring absolute vacuum stability within the SM 
\cite{NNLO_stability} although it was also shown \cite{bezrukov_rubio_shapo} that traditional Higgs inflation can be possible 
within a minimalistic framework even if the SM vacuum is not completely stable. 

Therefore, the UV completion of the Higgs potential which try to make connection between Higgs and inflationary physics, is 
required to have no just a global but local minima which can be realised by adding new interaction terms to the Lagrangian, such as 
$\lambda_2 (\phi^\star \phi)^3$. Indeed, the stability study of various types of polynomial Higgs potentials in the framework of 
simplified models has been performed by using functional renormalization group (RG) technique \cite{higgs_frg}. However, in the 
RG point of view the phase structure has not been modified significantly if the Higgs potential remains polynomial. Instead, a periodic 
self-interaction,
\beq
\label{sg_higgs}
V = u[\cos(\beta \sqrt{\phi^\star \phi})-1] = -\frac{u \beta^2}{2}  \vert \phi^\star \phi \vert + \frac{u \beta^4}{4}   (\phi^\star \phi)^2 + ...
\nonumber
\eeq
can influence more drastically the phase structure and the RG running of the couplings due to the periodicity which should be protected 
by the RG. The frequency $\beta$ can be chosen to keep the first minimum of the potential to be equal to that of the standard quartic 
Higgs potential, i.e., $\beta=\sqrt{2}\pi/v$. The inclusion of higher harmonics can shift the second minimum to take place at large fields.
There are various scenarios to identify the Higgs mass by the appropriate choice of $u$ (i) using the Taylor expansion of the (tree-level, 
bare) potential around the zero field where one finds $\mu^2 = -\hf u \beta^2$, (ii) including the RG running of the parameters in the 
co-called massive phase of the periodic model where the Taylor expansion is more justified and the mass parameter is given by the 
infra-red (IR) values $\mu^2 = -\hf u_{\mr{IR}} \beta_{\mr{IR}}^2$, (iii) by means of the soliton mass. 

With the same parametrisation of the field around the (first) minimum, the Lagrangian of the Higgs sector looks
similar to the usual one but with a periodic self-interaction term for $h(x)$. 
\bea
\label{sg_higgs_2}
{\cal L} &=& \hf \partial_\mu h \partial^\mu h - u \left[\cos\left(\frac{1}{\sqrt{2}} \beta |v+h(x)|\right)-1\right] \nn
&+& \left(M^2_W W^+_\mu W^{-\,\mu} + \hf M^2_Z Z_\mu Z^\mu \right) \left(1+2\frac{h}{v}+\frac{h^2}{v^2}\right).
\nonumber
\eea
A similar structure is found for the $O(N)$ extended version of the sine-Gordon model \cite{o(n)_sg} which was studied in $d=2$ 
dimensions and the dependence of its phase structure on $N$ was determined but it has not been investigated in higher dimensions. 
A rigorous functional RG study provides us a very powerful tool to attack such problems as it has been done for the pure $O(N)$ 
model in higher dimensions where the triviality of the model is re-examined \cite{o(n)_large_d}. The assumption of a periodic 
Higgs potential means infinitely many degenerate minima and requires a careful functional RG treatment.

Another goal of this work is to discuss some of the consequences of the renormalization of the pure sine-Gordon model to 
Higgs inflation and to the stability problem in the framework of the Higgs-Yukawa theory.

\section{Periodic axion potential}
\label{axion}
Our last example for applications of sine-Gordon type scalar field theory in higher dimensions is related to the axion. Constraints
from symmetry and renormalizability on the standard model QCD action allows to extend it by a CP violating term. However, experimental
data do not favour such an extension although the standard model Lagrangian is not CP symmetric, so, QCD could be CP violating as well.
Peccei and Quinn proposed a mechanism and introduced a new hypothetical scalar field with $U(1)$ symmetry in order to build up a CP 
conserving theory from a model with massive fermions coupled to a non-Abelian gauge field \cite{axion}. The axion appears as a phase of a 
Goldstone mode for a complex scalar $\Phi$ with a vacuum expectation value $<\Phi> = f e^{i\theta}$ corresponds to the spontaneous break 
down of the $U(1)$ symmetry at the scale $f$. Integrating over the QCD degrees of freedom one arrives at the following effective action
\beq
S = \int d^4 x \left(\frac{f^2}{2} \partial_\mu \theta \partial^\mu \theta + u [1-\cos(\theta)]\right) 
= \int d^4 x \left(\frac{1}{2} \partial_\mu \phi \partial^\mu \phi + u [1-\cos(\beta \phi)]\right), 
\nonumber
\eeq
with $\phi = f \theta$ and $\beta = 1/f$ where the periodic potential appears naturally and the rescaling of the field has been done by using 
the assumption that $f$ is independent of the spacetime. In Ref. \cite{axion_flattening} it was shown that the axion potential flattens out 
under RG transformations which were taken in the so called local potential approximation. Our goal here is to perform the functional RG study 
of the periodic scalar model, i.e., the sine-Gordon theory in $d=4$ beyond the local potential approximation in order to clarify the flattening 
of the axion potential.

\end{document}